\documentclass[10pt,journal,compsoc]{IEEEtran}
\ifCLASSOPTIONcompsoc
  \usepackage[nocompress]{cite}
\else
  \usepackage{cite}
\fi
\ifCLASSINFOpdf
  \usepackage[pdftex]{graphicx}
  \graphicspath{{../images/}}
  \DeclareGraphicsExtensions{.pdf,.jpeg,.png}
\else
\fi
\usepackage{amsmath}
\usepackage{algorithmic}

\usepackage{array}
\usepackage{url}

\usepackage[table]{xcolor}
\usepackage{booktabs}
\usepackage{multirow}
\usepackage{multicol}
\usepackage{siunitx}
\usepackage{comment}
\usepackage[acronym, nopostdot]{glossaries}
\glsdisablehyper
\newacronym{NaN}{NaN}{Not a Number}
\newacronym{PAU}{PAU}{Posit Arithmetic Unit}
\newacronym{ISA}{ISA}{Instruction Set Architecture}
\newacronym{GEMM}{GEMM}{General Matrix Multiplication}
\newacronym{NaR}{NaR}{Not-a-Real}
\newacronym{RISC}{RISC}{Reduced Instruction Set Computer}
\newacronym{FPU}{FPU}{Floating-Point Unit}
\newacronym{MAC}{MAC}{Multiply-Accumulate}
\newacronym{SoC}{SoC}{System on a Chip}
\newacronym{FPGA}{FPGA}{Field-Programmable Gate Array}
\newacronym{ASIC}{ASIC}{Application-Specific Integrated Circuit}
\newacronym{LUT}{LUT}{Lookup Table}
\newacronym{DNN}{DNN}{Deep Neural Network}
\newacronym{ALU}{ALU}{Arithmetic Logic Unit}
\newacronym{MSE}{MSE}{Mean Squared Error}
\newacronym{FF}{FF}{Flip-flop}
\usepackage{soul}
\usepackage{eso-pic}
\newcommand{\myCommentDMQ}[1]{#1}
\newcommand{\myCommentDMQbis}[1]{#1}
\newcommand{\positenv}[2]{Posit$\langle #1, #2 \rangle$}

\newcommand{\instbit}[1]{\mbox{\scriptsize #1}}
\newcommand{\instbitrange}[2]{~\instbit{#1} \hfill \instbit{#2}~}
\newcommand{\SWITCH}[1]{\STATE \textbf{switch} (#1)}
\newcommand{\ENDSWITCH}{\STATE \textbf{end switch}}
\newcommand{\CASE}[1]{\STATE \textbf{case} #1\textbf{:} \begin{ALC@g}}
\newcommand{\CASECOMMENT}[2]{\STATE \textbf{case} #1\textbf{:} \COMMENT{#2}\begin{ALC@g}}
\newcommand{\ENDCASE}{\end{ALC@g}}

\newcommand{\DEFAULT}{\STATE \textbf{default:} \begin{ALC@g}}
\newcommand{\DEFAULTCOMMENT}[1]{\STATE \textbf{default:} \COMMENT{#1}\begin{ALC@g}}
\newcommand{\ENDDEFAULT}{\end{ALC@g}}
\newcommand{\DEFAULTLINE}[1]{\STATE \textbf{default:} }

\begin{document}
%
\title{PERCIVAL: Open-Source Posit RISC-V Core with Quire Capability}
%
%
%
%

\author{David~Mallasén,
        Raul~Murillo,
        Alberto~A.~Del~Barrio,~\IEEEmembership{Senior~Member,~IEEE,}
        Guillermo~Botella,~\IEEEmembership{Senior~Member,~IEEE,}
        Luis~Piñuel
        and~Manuel~Prieto-Matias
\thanks{All authors are with the Department of Computer Architecture and Automation, Complutense University of Madrid, 28040 Madrid, Spain.\protect\\
E-mails: \{dmallase, ramuri01, abarriog, gbotella, lpinuel, mpmatias\}@ucm.es}
\thanks{Manuscript received -; revised -.}%
}

\AddToShipoutPictureBG*{%
  \AtPageUpperLeft{%
    \setlength\unitlength{1in}%
    \hspace*{\dimexpr0.5\paperwidth\relax}
    \makebox(0,-0.5)[c]{\footnotesize This article has been accepted for publication in a future issue of this journal, but has not been fully edited. Content may change prior to final publication.}%
}}
\AddToShipoutPictureBG*{%
  \AtPageUpperLeft{%
    \setlength\unitlength{1in}%
    \hspace*{\dimexpr0.5\paperwidth\relax}
    \makebox(0,-0.8)[c]{\footnotesize Citation information: DOI  10.1109/TETC.2022.3187199, IEEE Transactions on Emerging Topics in Computing}%
}}
\AddToShipoutPictureBG*{%
  \AtPageLowerLeft{%
    \setlength\unitlength{1in}%
    \hspace*{\dimexpr0.5\paperwidth\relax}
    \makebox(0,0.8)[c]{\footnotesize This work is licensed under a Creative Commons Attribution-NonCommercial-NoDerivatives 4.0 License.}%
}}
\AddToShipoutPictureBG*{%
  \AtPageLowerLeft{%
    \setlength\unitlength{1in}%
    \hspace*{\dimexpr0.5\paperwidth\relax}
    \makebox(0,0.5)[c]{\footnotesize For more information, see https://creativecommons.org/licenses/by-nc-nd/4.0/}%
}}

\IEEEtitleabstractindextext{%
\begin{abstract}
The posit representation for real numbers is an alternative to the ubiquitous IEEE 754 floating-point standard. In this work, we present PERCIVAL, an application-level posit RISC\nobreakdash-V core based on CVA6 that can execute all posit instructions, including the quire fused operations. This solves the obstacle encountered by previous works, which only included partial posit support or which had to emulate posits in software. In addition, Xposit, a RISC\nobreakdash-V extension for posit instructions is incorporated into LLVM. Therefore, PERCIVAL is the first work that integrates the complete posit instruction set in hardware. These elements allow for the native execution of posit instructions as well as the standard floating-point ones, further permitting the comparison of these representations. FPGA and ASIC synthesis show the hardware cost of implementing 32-bit posits and highlight the significant overhead of including a quire accumulator. However, results show that the quire enables a more accurate execution of dot products. In general matrix multiplications, the accuracy error is reduced up to 4 orders of magnitude. Furthermore, performance comparisons show that these accuracy improvements do not hinder their execution, as posits run as fast as single-precision floats and exhibit better timing than double-precision floats, thus potentially providing an alternative representation.
\end{abstract}

}

\maketitle

\IEEEdisplaynontitleabstractindextext

%
\IEEEpeerreviewmaketitle

\IEEEraisesectionheading{\section{Introduction}\label{sec:introduction}}

%
%
%
%


\IEEEPARstart{R}{epresenting} real numbers and executing arithmetic operations on them in a microprocessor presents unique challenges. When comparing with the simpler set of integers, working with reals introduces notions such as their precision. The representation of real numbers in virtually all computers for decades has followed the IEEE 754 standard for floating-point arithmetic~\cite{ieeecomputersociety2019IEEE}. However, this standard has some flaws such as rounding and reproducibility issues, signed zero, or excess of \gls{NaN} representations.

To face these challenges, alternative real number representations are proposed in the literature. Posits~\cite{gustafson2017Beating} are a promising substitute proposed in 2017 that provide compelling benefits. They deliver a good trade-off between dynamic range and accuracy, encounter fewer exceptions when operating, and have tapered precision. This means that numbers near $\pm 1$ have more precision, while very big and very small numbers have less. The posit standard includes fused operations, which can be used to compute a series of multiplications and accumulations without intermediate rounding. Furthermore, posits are consistent between implementations, as they use a single rounding scheme and include only two special cases: single 0 and $\pm\infty$. Therefore, they potentially simplify the hardware implementation~\cite{guntoro2020Next}. Nonetheless, posits are still under development, and it is still not clear whether they could completely replace IEEE floats~\cite{dedinechin2019Posits}.

Including \glspl{PAU} into cores in hardware is a crucial step to study the efficiency of this representation further. When designing such a core and its arithmetic operations, an important decision is which \gls{ISA} to implement. RISC\nobreakdash-V~\cite{waterman2014RISCV} is a promising open-source \gls{ISA} that is getting significant attraction both in academia and in industry. Thanks to its openness and flexibility, multiple RISC\nobreakdash-V cores have been developed targeting diverse purposes in recent years. In the case of studying the performance of posits, a core that can run application-level software is needed.

Some works have studied the use of posits by emulating their execution in software~\cite{murillo2020Deep, raposo2021Positnn, langroudi2021Alps}. However, this approach has the significant drawback of requiring excessive execution times, thus limiting the scalability of the applications.

To overcome these limitations, we propose to include native posit and quire support in hardware by leveraging a high-performance RISC\nobreakdash-V core. A comparison of four of the leading open-source application-class RISC\nobreakdash-V cores is studied in~\cite{dorflinger2021Comparative}, CVA6 among them. In this work, we have extended the datapath of the CVA6~\cite{zaruba2019Cost} RISC\nobreakdash-V core with a 32-bit \gls{PAU} with quire and a posit register file. Together with the Xposit compiler extension, this core allows the native hardware execution of high-level applications that leverage the posit number system.

Therefore, the main contributions of this paper are the following:
\begin{itemize}
    \item We present PERCIVAL, an oPEn-souRCe\footnote{\myCommentDMQ{\url{https://github.com/artecs-group/PERCIVAL}}} posIt risc-V core with quire cApabiLity based on the CVA6 that can execute all 32-bit posit instructions, including the quire fused operations.
    \item Compiler support for the Xposit RISC\nobreakdash-V extension in LLVM. This allows to easily embed posit instructions into a C program that can be run natively on PERCIVAL or any other core that implements these opcodes.
    \item To the best of our knowledge, the PERCIVAL core together with the Xposit extension is the first work that integrates in hardware \myCommentDMQbis{standard posit addition, subtraction, and multiplication together with quire fused operations. It also includes posit logarithmic-approximate hardware for division and square root operations. Furthermore, all comparison operations and conversions to and from integer numbers are also included in PERCIVAL.} 
    \item \gls{FPGA} and \gls{ASIC} synthesis results showcasing the resource-usage of posit arithmetic and quire capabilities on a RISC\nobreakdash-V CPU. These results are compared with the native IEEE 754 \gls{FPU} available in the CVA6 and with previous works.
    \item Accuracy and timing performance of posit numbers and IEEE 754 floats are compared on PERCIVAL using \gls{GEMM} and max-pooling benchmarks. Results show that 32-bit posits can be up to 4 orders of magnitude more accurate than 32-bit floats thanks to the quire register. Furthermore, this improvement does not imply a trade-off in execution time, as they can perform as fast as 32-bit floats, and thus execute faster than 64-bit floats.
\end{itemize}

The rest of the paper is organized as follows: Section~\ref{sec:background} introduces the necessary background about the posit format, the RISC\nobreakdash-V \gls{ISA} and the CVA6 RISC\nobreakdash-V core. Related works from the literature are surveyed in Section~\ref{sec:related_work}, both as standalone \glspl{PAU} and at the core level. In Section~\ref{sec:posit_core} the PERCIVAL posit core is described and in Section~\ref{sec:compiler_support} the necessary compiler support for the Xposit RISC\nobreakdash-V extension is introduced. The \gls{FPGA} and \gls{ASIC} synthesis results of the core are presented, as well as compared with other implementations, in Section~\ref{sec:synthesis_results}. Subsequently, in Section~\ref{sec:posit_vs_ieee} posits and IEEE 754 floats are compared regarding accuracy and timing performance. Finally, Section~\ref{sec:conclusions} concludes this work.

\section{Background} \label{sec:background}

\subsection{Posit Format} \label{sec:posit_format}

Posit numbers~\cite{gustafson2017Beating} were introduced in 2017 as an alternative to the predominant IEEE 754 floating-point standard to represent and operate with real numbers. Posits provide reproducible results across platforms and few special cases. Furthermore, they do not support overflow or underflow, which reduces the complexity of exception handling.

A posit number configuration is defined using two parameters as \positenv{n}{es}, where $n$ is the total bit-width, and $es$ is the maximum bit-width of the exponent. Although in literature~\cite{dedinechin2019Posits, murillo2021PLAM, murillo2020Deep} the most widespread posit formats have been \positenv{8}{0}, \positenv{16}{1} and \positenv{32}{2}, in the latest Posit Standard 4.12  Draft~\cite{positworkinggroup2021Posit}, the value of $es$ is fixed to 2. This has the advantage of simplifying the hardware design and facilitates the conversion between different posit sizes.

\begin{figure}
    \centering
    \includegraphics[width=0.95\columnwidth]{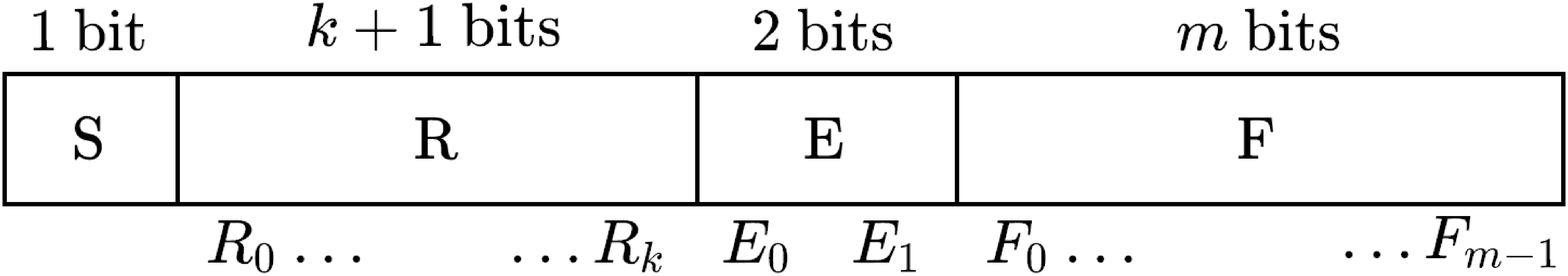}
    \caption{Posit format with sign, regime, exponent and fraction fields.}
    \label{fig:posit_format}
\end{figure}

Posits only distinguish two special cases: zero and \gls{NaR}, which are represented as $\texttt{0}\cdots\texttt{0}$ and $\texttt{10}\cdots\texttt{0}$ respectively. The rest of the representations are composed of four fields as shown in Figure~\ref{fig:posit_format}:
\begin{itemize}
    \item The sign bit S;
    \item The variable-length regime field R, consisting of $k$ bits equal to $R_0$ followed by $\overline{R_0}$ or the end of the posit. This field encodes a scaling factor $r$ given by Equation~(\ref{eq:regime_value});
    \item The exponent E, consisting of at most $es$ bits, which encodes an integer unbiased value $e$. If any of its bits are located after the least significant bit of the posit, that bit will have value \texttt{0};
    \item The variable-length fraction field F, formed by the remaining $m$ bits. Its value $0 \leq f < 1$ is given by dividing the unsigned integer F by $2^m$.
\end{itemize}
\begin{equation} \label{eq:regime_value}
    r = \left\{
	\begin{array}{ll}
		-k & \mbox{if } R_0 = 0 \\
		k-1 & \mbox{if } R_0 = 1
	\end{array}
	\right.
\end{equation}

The real value $p$ of a generic posit is given by Equation~(\ref{eq:posit_value}). The main differences with the IEEE 754 floating-point format are the existence of the regime field, the use of an unbiased exponent, and the value of the fraction hidden bit. Usually, in floating-point arithmetic, the hidden bit is considered to be $1$. However, in the latest representation of posits, it is considered to be $1$ if the number is positive, or $-2$ if the number is negative. This simplifies the decoding stage of the posit representation~\cite{guntoro2020Next},~\myCommentDMQ{{\mbox{\cite{murillo2022Comparing}}}}.
\begin{equation} \label{eq:posit_value}
    p = ((1 - 3s) + f)\times 2^{(1-2s)\times(4r + e + s)}
\end{equation}

In posit arithmetic, \gls{NaR} has a unique representation that maps to the most negative 2's complement signed integer. Consequently, if used in comparison operations, it results in less than all other posits and equal to itself. Moreover, the rest of the posit values follow the same ordering as their corresponding bit representations. These characteristics allow posit numbers to be compared as if they were 2's complement signed integers, eliminating additional hardware for posit comparison operations.

The variable-length regime field acts as a long-range dynamic exponent, as can be seen in Equation~(\ref{eq:posit_value}), where it is multiplied by 4 or, equivalently, shifted left by the two exponent bits. Since it is a dynamic field, it can occupy more bits to represent larger numbers or leave more bits to the fraction field when looking for accuracy in the neighborhoods of $\pm 1$. However, detecting these variable-sized fields adds some hardware overhead.

As an example, let \texttt{11101010} be the binary encoding of a Posit8, i.e. a \positenv{8}{2} according to the latest Posit Standard 4.12 Draft~\cite{positworkinggroup2021Posit}. The first bit $s=\texttt{1}$ indicates a negative number. The regime field \texttt{110} gives $k=2$ and therefore $r=1$. The next two bits \texttt{10} represent the exponent $e=2$. Finally, the remaining $m=2$ bits, \texttt{10}, encode a fraction value of $f=2/2^2=0.5$. Hence, from~(\ref{eq:posit_value}) we conclude that $\texttt{11101010}\equiv (-2 + 0.5)\times 2^{-(4+2+1)} = -0.01171875.$

In addition to the standard representation, posits include fused operations using the \emph{quire}, a $16n$-bit fixed-point 2's complement register, where $n$ is the posit bit-width. This allows to execute up to $2^{31} - 1$ \gls{MAC} operations without intermediate rounding or accuracy loss. The quire can represent either \gls{NaR}, similarly to regular posits, or the value given by $2^{16-8n}$ times the 2's complement signed integer represented by the $16n$ concatenated bits.

\subsection{RISC-V ISA}

The open-source RISC\nobreakdash-V \gls{ISA}~\cite{waterman2014RISCV} emanates from the ideas of \glspl{RISC}. It is structured as a base integer \gls{ISA} plus a set of optional standard and non-standard extensions to customize and specialize the final set of instructions. There are two main base integer \glspl{ISA}, RV32I and RV64I, that establish the \myCommentDMQ{user} address spaces as 32-bit or 64-bit respectively.

The RISC\nobreakdash-V general standard extensions include, among others, functionality for integer multiply/divide (M), atomic memory operations (A) and single- (F) and double-precision (D) floating-point arithmetic following the IEEE 754 standard. \myCommentDMQ{This set of general-purpose standard extensions IMAFD, together with the instruction-fetch fence (Zifencei), and the control and status register (Zicsr), form the general-purpose G abbreviation.} In general, following the \gls{RISC} principles, all extensions have fixed-length 32-bit instructions. However, there is also a compressed instruction extension (C) that provides 16-bit instructions.

Expanding the RISC\nobreakdash-V \gls{ISA} with specialized extensions is supported by the standard to allow for customized accelerators. Non-standard extensions can be added to the encoding space leveraging the four major opcodes reserved for custom extensions. A proposal of the changes that should be made to the F standard extension in order to have a 32-bit posit RISC\nobreakdash-V extension is described in~\cite{gustafson2018RISCV}.

\subsection{CVA6}

The CVA6~\cite{zaruba2019Cost} (formerly known as Ariane) is a 6-stage, in-order, single-issue CPU which implements the RV64GC RISC\nobreakdash-V standard. The core implements three privilege levels and can run a Linux operating system. The primary goal of its micro-architecture is to reduce the critical path length. It was developed initially as part of the PULP ecosystem, but it is currently maintained by the OpenHW Group, which is developing a complete, industrial-grade pre-silicon verification. CVA6 is written in SystemVerilog and is licensed under an open-source Solderpad Hardware License. 

As execution units in the datapath it includes an integer ALU, a multiply/divide unit and an IEEE 754 \gls{FPU} \cite{mach2021FPnew}. This \gls{FPU} claims to be IEEE 754-2008 compliant, except for some issues in the division and square root operations. For the sake of comparison, it is important that the \gls{FPU} is IEEE 754 compliant instead of being limited to normal floats only, since in theory, posit hardware is slightly more expensive than floating-point hardware that does not take into account subnormal numbers~\cite{guntoro2020Next}.

\section{Related Work} \label{sec:related_work}
There has been a great deal of interest in hardware implementations of posit arithmetic since its first appearance. Standalone \glspl{PAU} with different degrees of capabilities or basic posit functional units have been described in the literature~\cite{chaurasiya2018Parameterized,jaiswal2019PACoGen,murillo2020Customized,murillo2021PLAM}. These units provide the building blocks to execute posit arithmetic. However, they do not allow themselves to execute whole posit algorithms.

Recently, some works adding partial posit support to RISC\nobreakdash-V cores have been presented. CLARINET~\cite{sharma2021CLARINET} incorporates the quire into a RV64GC 5-stage in-order core. However, not all posit capabilities are included in this work. Most operations are performed in IEEE floating-point format, and the values are converted to posit when using the quire. The only posit functionalities added to the core are fused \gls{MAC} with quire, fused divide and accumulate with quire and conversion instructions.

PERC~\cite{arunkumar2020PERC} integrates a \gls{PAU} into the Rocket Chip generator, replacing the 32 and 64-bit \gls{FPU}. However, this work does not include quire support, as it is constrained by the F and D RISC\nobreakdash-V extensions for IEEE-754 floating-point numbers. More recently, PERI~\cite{tiwari2021PERI} added a tightly coupled \gls{PAU} into the SHAKTI C-class core, a 5-stage in-order RV32IMAFC core. This proposal also does not include quire support as it reuses the F extension instructions. Nonetheless, it allows dynamic switching between es=2 and es=3 posits. In~\cite{ciocirlan2021Accuracy} authors include a \gls{PAU} named POSAR into a RISC\nobreakdash-V Rocket Chip core. Again, this proposal does not include quire support and replaces the \gls{FPU} present in Rocket Chip to reuse the floating-point instructions.

A different approach is taken in~\cite{cococcioni2021Lightweight}, where authors use the posit representation as a way to store IEEE floats in memory with a lower bit-width while performing the computations using the IEEE \gls{FPU}. For this purpose they include a light posit processing unit into the CVA6 core that converts between 8 or 16-bit posits and 32-bit IEEE floats. They also develop an extension of the RISC\nobreakdash-V ISA to include these conversion instructions.

\section{PERCIVAL Posit Core} \label{sec:posit_core}

In this work, we have integrated full posit capabilities, including quire and fused operations, into an application-level RISC\nobreakdash-V core. In addition to the design of the functional units that execute the posit and quire operations, the novelty of our design is that it is fully compatible both at the software and hardware level with the F and D RISC\nobreakdash-V extensions. Therefore, both posit and IEEE floating-point numbers can be used simultaneously on the same core. This is the first work that integrates \myCommentDMQ{practically all of the posit and quire operations specified in the posit standard} into a core, to the best of our knowledge.

\subsection{PAU Design}

\begin{figure*}
    \centering
    \includegraphics[width=0.9\textwidth]{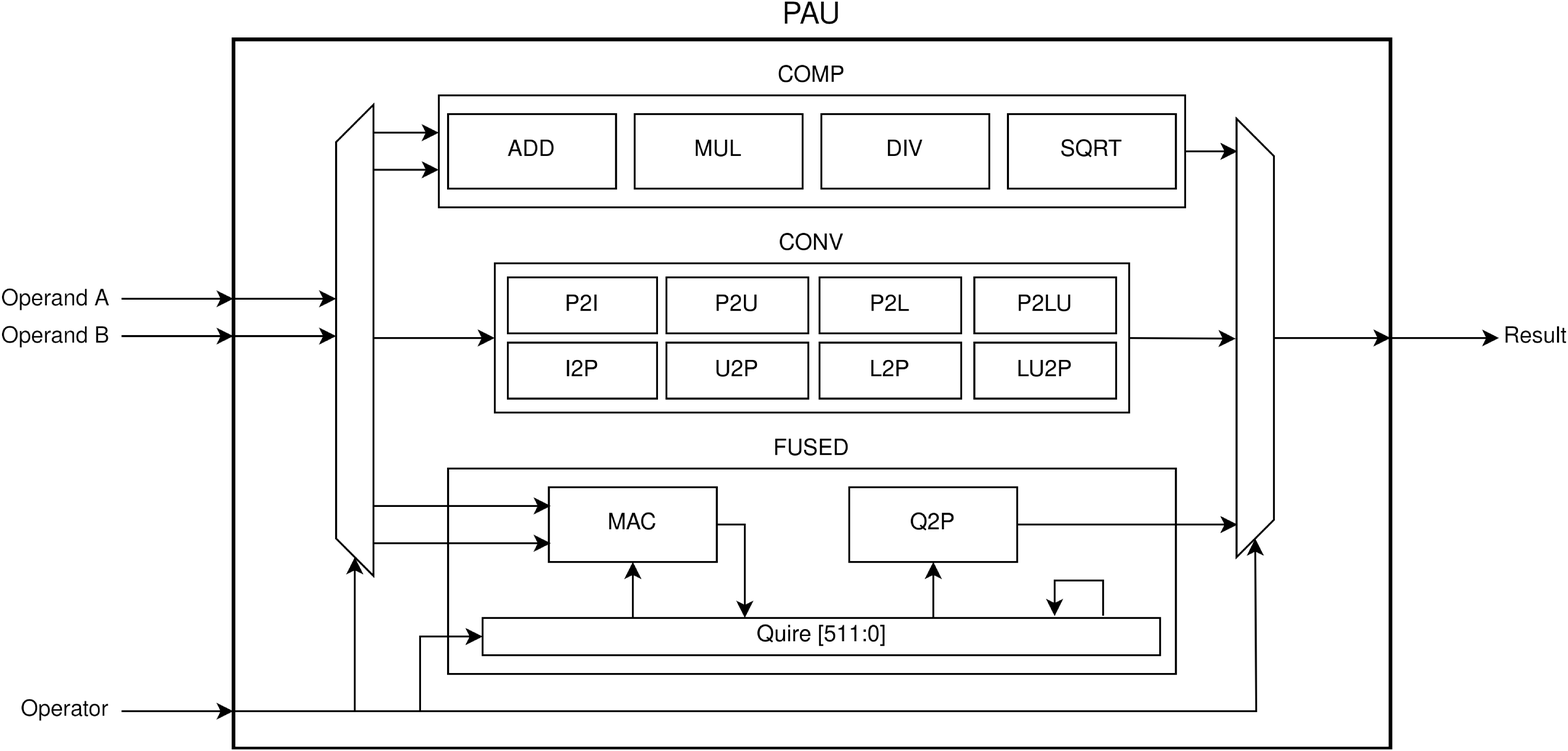}
    \caption{Internal structure of the proposed \acrfull{PAU}.}
    \label{fig:pau}
\end{figure*}

The \acrfull{PAU} is in charge of executing most posit operations and also contains the quire register, as shown in Figure~\ref{fig:pau}. Posit comparisons are executed in the integer \gls{ALU}. As mentioned above, this is one of the benefits of the posit representation. 

\myCommentDMQ{When designing the micro-architecture of the {\gls{PAU}}, our objective was to achieve a similar latency and throughput as the {\gls{FPU}} operations, to obtain fair comparisons. The throughput is limited, as there is no pipeline in the {\gls{FPU}} nor the {\gls{PAU}}. Nevertheless, all of the operations are multi-cycle. The latency of the {\gls{PAU}} units is the following:}
\begin{itemize}
    \item \myCommentDMQ{PADD, PSUB, QMADD, and QMSUB: 2 cycles.}
    \item \myCommentDMQ{PMUL, PDIV, PSQRT, and QROUND: 1 cycle.}
\end{itemize}
\myCommentDMQ{All other operations have no latency, i.e. they output their result at the next clock cycle after receiving the inputs.
}

\myCommentDMQ{As a comparison, the 32-bit FADD, FSUB, FMADD, FMSUB, and FMUL instructions in the {\gls{FPU}} have a latency of 2 clock cycles, but the 64-bit analogous instructions have a latency of 3 cycles. It is noteworthy that the comparisons in the {\gls{FPU}} have a latency of 1, while the posit comparisons that reuse the integer hardware have no latency. Conversions to and from integer values also take an extra clock cycle in the {\gls{FPU}}.
}

Depending on the operation, the input operands are directed to the corresponding posit unit and the result is forwarded as an output of the \gls{PAU}. There are three main blocks: computational operations (COMP), conversion operations (CONV), and operations that make use of the quire register (FUSED) (Figure~\ref{fig:pau}).

Regarding COMP, the ADD unit is used both for addition and subtraction, calculating the two's complement of the second operand when subtracting. In this group, all the modules use both operands except the square root, which uses only operand A. In addition, the operands and the result correspond to the posit register file.

It must be noted that the posit division and square root units are approximate, as this type of arithmetic simplifies the designs and thus reduces the hardware cost of the system. \myCommentDMQbis{They are logarithm-approximate units based on Mitchell’s Approximate Log Multipliers and our previous work}~\cite{murillo2021PLAM}. \myCommentDMQbis{These units have been demonstrated to have a maximum relative error of $11.11\%$, and have less impact on area/performance than the exact hardware operators.} On the other hand, exact division and square root algorithms could be implemented in software leveraging the MAC unit, thus eliminating the need for dedicated hardware. However, this is out of the scope of this work.

In the CONV group, only operand A is used for conversions. Depending on the operation, the input data and the result belong to the posit or the integer register file.

The quire register is the most singular addition to this number format. According to the posit standard, it must be an architectural register accessible by the programmer that is also allowed to be dumped into memory. However, being so wide, the cost of including this functionality into the core's datapath could be too high for the benefits it would add. In the vast majority of cases, the quire is used as an accumulator to avoid overflows in the \gls{MAC} operations, and this does not require quire load and store operations. Instead, we can initialize the quire to zero (QCLR.S), negate it if needed (QNEG.S), accumulate the partial products in it without rounding or storing in memory (QMADD.S and QMSUB.S), and, when the whole operation is finished, round and output the result (QROUND.S). The necessary support for all of these operations related to the quire is included in our proposal (see Table~\ref{tab:xposit_instructions} below). The hardware cost of including the quire as an internal register in the \gls{PAU} is studied in Section~\ref{sec:synthesis_results}.

\subsection{Core Integration}

The proposed \gls{PAU} has been integrated into the CVA6 RV64GC core while maintaining the compatibility with all existing extensions, including single- and double-precision floating point. Moreover, since we work with Posit32 numbers, i.e. \positenv{32}{2}, the core adds a 32-bit posit register file in addition to the integer and floating-point registers.

The instruction decoder has been extended to support posit instructions. The inner workings of the decoder are described in Figure~\ref{alg:decoder}. As part of the decoding process, each posit instruction selects from which register file it must obtain its operands and to which register file it must forward its result.

\begin{figure}
\begin{algorithmic}
    \REQUIRE Instruction to decode \texttt{instr}.
    \ENSURE Scoreboard entry \texttt{sc\_instr} which contains the operation \texttt{op} and the destination functional unit \texttt{fu}.
    \SWITCH {\texttt{instr.opcode}}
        \STATE $\dots$
        \CASE {\texttt{POSIT}}
            \SWITCH {\texttt{instr.func3}}
                \CASECOMMENT {\texttt{000}}{Computational posit instruction}
                    \SWITCH {\texttt{instr.func5}}
                        \CASECOMMENT {\texttt{00000}}{PAU instruction}
                            \STATE \texttt{sc\_instr.fu} = \texttt{PAU}
                            \STATE \texttt{sc\_instr.op} = \texttt{PADD}
                        \ENDCASE
                        \STATE $\dots$
                        \CASECOMMENT {\texttt{00100}}{ALU instruction}
                            \STATE \texttt{sc\_instr.fu} = \texttt{ALU}
                            \STATE \texttt{sc\_instr.op} = \texttt{PMIN}
                        \ENDCASE
                        \STATE $\dots$
                    \ENDSWITCH
                \ENDCASE
                \CASECOMMENT {\texttt{001}}{Posit load instruction}
                    \STATE \texttt{sc\_instr.fu} = \texttt{LOAD}
                    \STATE \texttt{sc\_instr.op} = \texttt{PLW}
                \ENDCASE
                \CASECOMMENT {\texttt{011}}{Posit store instruction}
                    \STATE \texttt{sc\_instr.fu} = \texttt{STORE}
                    \STATE \texttt{sc\_instr.op} = \texttt{PSW}
                \ENDCASE
            \ENDSWITCH
        \ENDCASE
        \STATE $\dots$
        \DEFAULTCOMMENT{Instruction not decoded in any switch/case}
            \STATE \texttt{illegal\_instr} = \TRUE
        \ENDDEFAULT
    \ENDSWITCH
\end{algorithmic}
\caption{Pseudocode describing the decoding of posit instructions.}
\label{alg:decoder}
\end{figure}

The CVA6 core uses scoreboarding for dynamically scheduled instructions and allows out-of-order write-back of each functional unit. The scoreboard tracks which instructions are issued, their functional unit and in which register they will write back to. Our design has enlarged the scoreboard to include posit registers and instructions. In this manner, we can discern whether the input data of posit operations are retrieved from a register or forwarded directly as a result of a previous operation.

As mentioned in Section~\ref{sec:posit_format}, posit numbers have the benefit of being able to reuse the comparison hardware of 2's complement signed integers. Therefore, the integer \gls{ALU} has also been extended to accept posit operands and to be able to forward the result of these instructions with minimal hardware overhead. Furthermore, the \gls{PAU} has been integrated into the execution phase of the processor in parallel to the \gls{ALU} and the \gls{FPU}, connecting the issue module with the aforementioned scoreboard. Finally, the complete datapath has been adapted to include the posit signals and all necessary additional interconnections.

\section{Compiler Support: Xposit extension} \label{sec:compiler_support}

The assembly output of a RISC\nobreakdash-V compiler when processing programs that use floating-point arithmetic includes instructions from the corresponding F and D extensions. To produce a similar output but targeting posit numbers, a new extension must be introduced that translates posit instructions and posit operators to binary code. Therefore, in this section, the Xposit RISC\nobreakdash-V extension targeting posit arithmetic is presented. As part of this work, Xposit has been integrated into LLVM 12 backend~\cite{lattner2004LLVM} to allow the compilation of high-level applications.

\myCommentDMQbis{
This modified version of LLVM can compile C code. However, posit instructions must be written from the assembly level, as there is currently no support for writing posit or quire operations directly in C. Therefore, previous codes can be reused in PERCIVAL, and only the computational kernels have to be manually written in assembly. An example of this is shown in Section~\ref{sec:posit_vs_ieee}.
}

The posit instruction set follows the structure of the F RISC\nobreakdash-V standard extension for single-precision floating point~\cite{waterman2019RISCV}. This Xposit extension mostly follows the adaptation to the posit format proposed in \cite{gustafson2018RISCV}. The differences with this proposal are the following:
\begin{itemize}
    \item We include 32 posit registers \texttt{p0-31} as in the F standard extension.
    \item Similarly to the integer operations in CVA6, there is no flag signaling division by zero.
    \item We do not include the possibility of loading and storing the quire in memory.
\end{itemize}

The Xposit extension uses the 0001011 opcode ({\em custom\nobreakdash-0}), occupying the space indicated in Table~\ref{tab:opcodemap} as POSIT. If more operations were needed in the future, especially posit load and store instructions of other word lengths, the 0101011, 1011011, and 1111011 opcodes ({\em custom\nobreakdash-1,2,3}) could be leveraged. In this way, a similar approach as the F and D RISC\nobreakdash-V extensions could be followed, which utilize the OP\nobreakdash-FP, LOAD\nobreakdash-FP and STORE\nobreakdash-FP opcodes.

\definecolor{gray}{RGB}{180,180,180}
\definecolor{lightgray}{gray}{0.85}
\begin{table*}
\caption{RISC\nobreakdash-V base opcode map + POSIT extension; inst[1:0]=11}
\label{tab:opcodemap}
\centering
{\footnotesize
\setlength{\tabcolsep}{4pt}
\begin{tabular}{|r|c|c|c|c|c|c|c|c|}
  \hline
  inst[4:2] & 000    & 001      & 010            & 011      & 100    & 101            & 110                  & \cellcolor{gray}111 \\ \cline{1-1}
  inst[6:5] &        &          &                &          &        &                &                      & \cellcolor{gray}($>32b$)  \\ \hline
         00 & LOAD   & LOAD-FP  & \cellcolor{lightgray} POSIT & MISC-MEM & OP-IMM & AUIPC          & OP-IMM-32            & \cellcolor{gray} $48b$\\ \hline
         01 & STORE  & STORE-FP & {\em custom-1} & AMO      & OP     & LUI            & OP-32                & \cellcolor{gray} $64b$ \\ \hline
         10 & MADD   & MSUB     & NMSUB          & NMADD    & OP-FP  & {\em reserved} & {\em custom-2/rv128} & \cellcolor{gray} $48b$\\ \hline
         11 & BRANCH & JALR     & {\em reserved} & JAL      & SYSTEM & {reserved} & {\em custom-3/rv128} & \cellcolor{gray} $\geq80b$\\ \hline

 \end{tabular}
}
\end{table*}

\begin{figure*}
    \centering
    \includegraphics[width=0.7\textwidth]{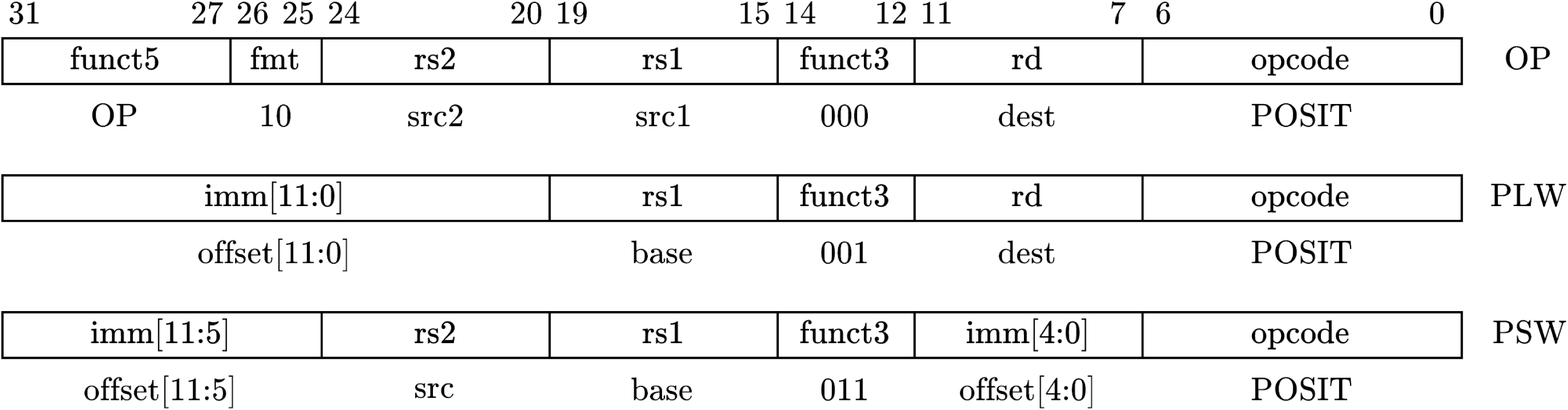}
    \caption{Internal structure and fields of Xposit instructions.}
    \label{fig:xposit_instruction_structure}
\end{figure*}

The format and fields of the Xposit instructions are described in Figure~\ref{fig:xposit_instruction_structure}. Posit load and store use the same base+offset addressing as the corresponding floating-point instructions, with the base address in register {\em rs1} and a signed 12-bit byte offset. Thus, the PLW instruction loads a posit value from memory to the {\em rd} posit register and the PSW instruction stores a posit value from the {\em rs2} posit register to memory. The rest of the Xposit operations keep the POSIT opcode and differ from the previous instructions by the {\em funct3} field. Finally, it must be noted that the {\em fmt} field is fixed to \texttt{01} indicating that the instructions are for single-precision (32-bit) posits. The complete instruction set of the proposed Xposit RISC\nobreakdash-V extension is detailed in Table~\ref{tab:xposit_instructions}.

\begin{table*}
\caption{Instruction set of the proposed XPosit RISC\nobreakdash-V extension.}
\label{tab:xposit_instructions}
\centering
\begin{tabular}{p{0in}p{0.4in}p{0.05in}p{0.05in}p{0.05in}p{0.05in}p{0.4in}p{0.6in}p{0.4in}p{0.6in}p{0.7in}l}
& & & & & & & & & & \\
&
\multicolumn{1}{l}{\instbit{31}} &
\multicolumn{1}{r}{\instbit{27}} &
\instbit{26} &
\instbit{25} &
\multicolumn{1}{l}{\instbit{24}} &
\multicolumn{1}{r}{\instbit{20}} &
\instbitrange{19}{15} &
\instbitrange{14}{12} &
\instbitrange{11}{7} &
\instbitrange{6}{0} \\
\cline{2-11}

&
\multicolumn{6}{|c|}{imm[11:0]} &
\multicolumn{1}{c|}{rs1} &
\multicolumn{1}{c|}{001} &
\multicolumn{1}{c|}{rd} &
\multicolumn{1}{c|}{00001011} & PLW \\
\cline{2-11}
  
&
\multicolumn{4}{|c|}{imm[11:5]} &
\multicolumn{2}{c|}{rs2} &
\multicolumn{1}{c|}{rs1} &
\multicolumn{1}{c|}{011} &
\multicolumn{1}{c|}{imm[4:0]} &
\multicolumn{1}{c|}{00001011} & PSW \\
\cline{2-11}

&
\multicolumn{2}{|c|}{00000} &
\multicolumn{2}{c|}{10} &
\multicolumn{2}{c|}{rs2} &
\multicolumn{1}{c|}{rs1} &
\multicolumn{1}{c|}{000} &
\multicolumn{1}{c|}{rd} &
\multicolumn{1}{c|}{00001011} & PADD.S \\
\cline{2-11}

&
\multicolumn{2}{|c|}{00001} &
\multicolumn{2}{c|}{10} &
\multicolumn{2}{c|}{rs2} &
\multicolumn{1}{c|}{rs1} &
\multicolumn{1}{c|}{000} &
\multicolumn{1}{c|}{rd} &
\multicolumn{1}{c|}{00001011} & PSUB.S \\
\cline{2-11}

&
\multicolumn{2}{|c|}{00010} &
\multicolumn{2}{c|}{10} &
\multicolumn{2}{c|}{rs2} &
\multicolumn{1}{c|}{rs1} &
\multicolumn{1}{c|}{000} &
\multicolumn{1}{c|}{rd} &
\multicolumn{1}{c|}{00001011} & PMUL.S \\
\cline{2-11}

&
\multicolumn{2}{|c|}{00011} &
\multicolumn{2}{c|}{10} &
\multicolumn{2}{c|}{rs2} &
\multicolumn{1}{c|}{rs1} &
\multicolumn{1}{c|}{000} &
\multicolumn{1}{c|}{rd} &
\multicolumn{1}{c|}{00001011} & PDIV.S \\
\cline{2-11}

&
\multicolumn{2}{|c|}{00100} &
\multicolumn{2}{c|}{10} &
\multicolumn{2}{c|}{rs2} &
\multicolumn{1}{c|}{rs1} &
\multicolumn{1}{c|}{000} &
\multicolumn{1}{c|}{rd} &
\multicolumn{1}{c|}{00001011} & PMIN.S \\
\cline{2-11}

&
\multicolumn{2}{|c|}{00101} &
\multicolumn{2}{c|}{10} &
\multicolumn{2}{c|}{rs2} &
\multicolumn{1}{c|}{rs1} &
\multicolumn{1}{c|}{000} &
\multicolumn{1}{c|}{rd} &
\multicolumn{1}{c|}{00001011} & PMAX.S \\
\cline{2-11}

&
\multicolumn{2}{|c|}{00110} &
\multicolumn{2}{c|}{10} &
\multicolumn{2}{c|}{00000} &
\multicolumn{1}{c|}{rs1} &
\multicolumn{1}{c|}{000} &
\multicolumn{1}{c|}{rd} &
\multicolumn{1}{c|}{00001011} & PSQRT.S \\
\cline{2-11}

&
\multicolumn{2}{|c|}{00111} &
\multicolumn{2}{c|}{10} &
\multicolumn{2}{c|}{rs2} &
\multicolumn{1}{c|}{rs1} &
\multicolumn{1}{c|}{000} &
\multicolumn{1}{c|}{00000} &
\multicolumn{1}{c|}{00001011} & QMADD.S \\
\cline{2-11}

&
\multicolumn{2}{|c|}{01000} &
\multicolumn{2}{c|}{10} &
\multicolumn{2}{c|}{rs2} &
\multicolumn{1}{c|}{rs1} &
\multicolumn{1}{c|}{000} &
\multicolumn{1}{c|}{00000} &
\multicolumn{1}{c|}{00001011} & QMSUB.S \\
\cline{2-11}

&
\multicolumn{2}{|c|}{01001} &
\multicolumn{2}{c|}{10} &
\multicolumn{2}{c|}{00000} &
\multicolumn{1}{c|}{00000} &
\multicolumn{1}{c|}{000} &
\multicolumn{1}{c|}{00000} &
\multicolumn{1}{c|}{00001011} & QCLR.S \\
\cline{2-11}

&
\multicolumn{2}{|c|}{01010} &
\multicolumn{2}{c|}{10} &
\multicolumn{2}{c|}{00000} &
\multicolumn{1}{c|}{00000} &
\multicolumn{1}{c|}{000} &
\multicolumn{1}{c|}{00000} &
\multicolumn{1}{c|}{00001011} & QNEG.S \\
\cline{2-11}

&
\multicolumn{2}{|c|}{01011} &
\multicolumn{2}{c|}{10} &
\multicolumn{2}{c|}{00000} &
\multicolumn{1}{c|}{00000} &
\multicolumn{1}{c|}{000} &
\multicolumn{1}{c|}{rd} &
\multicolumn{1}{c|}{00001011} & QROUND.S \\
\cline{2-11}

&
\multicolumn{2}{|c|}{01100} &
\multicolumn{2}{c|}{10} &
\multicolumn{2}{c|}{00000} &
\multicolumn{1}{c|}{rs1} &
\multicolumn{1}{c|}{000} &
\multicolumn{1}{c|}{rd} &
\multicolumn{1}{c|}{00001011} & PCVT.W.S \\
\cline{2-11}

&
\multicolumn{2}{|c|}{01101} &
\multicolumn{2}{c|}{10} &
\multicolumn{2}{c|}{00000} &
\multicolumn{1}{c|}{rs1} &
\multicolumn{1}{c|}{000} &
\multicolumn{1}{c|}{rd} &
\multicolumn{1}{c|}{00001011} & PCVT.WU.S \\
\cline{2-11}

&
\multicolumn{2}{|c|}{01110} &
\multicolumn{2}{c|}{10} &
\multicolumn{2}{c|}{00000} &
\multicolumn{1}{c|}{rs1} &
\multicolumn{1}{c|}{000} &
\multicolumn{1}{c|}{rd} &
\multicolumn{1}{c|}{00001011} & PCVT.L.S \\
\cline{2-11}

&
\multicolumn{2}{|c|}{01111} &
\multicolumn{2}{c|}{10} &
\multicolumn{2}{c|}{00000} &
\multicolumn{1}{c|}{rs1} &
\multicolumn{1}{c|}{000} &
\multicolumn{1}{c|}{rd} &
\multicolumn{1}{c|}{00001011} & PCVT.LU.S \\
\cline{2-11}

&
\multicolumn{2}{|c|}{10000} &
\multicolumn{2}{c|}{10} &
\multicolumn{2}{c|}{00000} &
\multicolumn{1}{c|}{rs1} &
\multicolumn{1}{c|}{000} &
\multicolumn{1}{c|}{rd} &
\multicolumn{1}{c|}{00001011} & PCVT.S.W \\
\cline{2-11}

&
\multicolumn{2}{|c|}{10001} &
\multicolumn{2}{c|}{10} &
\multicolumn{2}{c|}{00000} &
\multicolumn{1}{c|}{rs1} &
\multicolumn{1}{c|}{000} &
\multicolumn{1}{c|}{rd} &
\multicolumn{1}{c|}{00001011} & PCVT.S.WU \\
\cline{2-11}

&
\multicolumn{2}{|c|}{10010} &
\multicolumn{2}{c|}{10} &
\multicolumn{2}{c|}{00000} &
\multicolumn{1}{c|}{rs1} &
\multicolumn{1}{c|}{000} &
\multicolumn{1}{c|}{rd} &
\multicolumn{1}{c|}{00001011} & PCVT.S.L \\
\cline{2-11}

&
\multicolumn{2}{|c|}{10011} &
\multicolumn{2}{c|}{10} &
\multicolumn{2}{c|}{00000} &
\multicolumn{1}{c|}{rs1} &
\multicolumn{1}{c|}{000} &
\multicolumn{1}{c|}{rd} &
\multicolumn{1}{c|}{00001011} & PCVT.S.LU \\
\cline{2-11}

&
\multicolumn{2}{|c|}{10100} &
\multicolumn{2}{c|}{10} &
\multicolumn{2}{c|}{rs2} &
\multicolumn{1}{c|}{rs1} &
\multicolumn{1}{c|}{000} &
\multicolumn{1}{c|}{rd} &
\multicolumn{1}{c|}{00001011} & PSGNJ.S \\
\cline{2-11}

&
\multicolumn{2}{|c|}{10101} &
\multicolumn{2}{c|}{10} &
\multicolumn{2}{c|}{rs2} &
\multicolumn{1}{c|}{rs1} &
\multicolumn{1}{c|}{000} &
\multicolumn{1}{c|}{rd} &
\multicolumn{1}{c|}{00001011} & PSGNJN.S \\
\cline{2-11}

&
\multicolumn{2}{|c|}{10110} &
\multicolumn{2}{c|}{10} &
\multicolumn{2}{c|}{rs2} &
\multicolumn{1}{c|}{rs1} &
\multicolumn{1}{c|}{000} &
\multicolumn{1}{c|}{rd} &
\multicolumn{1}{c|}{00001011} & PSGNJX.S \\
\cline{2-11}

&
\multicolumn{2}{|c|}{10111} &
\multicolumn{2}{c|}{10} &
\multicolumn{2}{c|}{00000} &
\multicolumn{1}{c|}{rs1} &
\multicolumn{1}{c|}{000} &
\multicolumn{1}{c|}{rd} &
\multicolumn{1}{c|}{00001011} & PMV.X.W \\
\cline{2-11}

&
\multicolumn{2}{|c|}{11000} &
\multicolumn{2}{c|}{10} &
\multicolumn{2}{c|}{00000} &
\multicolumn{1}{c|}{rs1} &
\multicolumn{1}{c|}{000} &
\multicolumn{1}{c|}{rd} &
\multicolumn{1}{c|}{00001011} & PMV.W.X \\
\cline{2-11}

&
\multicolumn{2}{|c|}{11001} &
\multicolumn{2}{c|}{10} &
\multicolumn{2}{c|}{rs2} &
\multicolumn{1}{c|}{rs1} &
\multicolumn{1}{c|}{000} &
\multicolumn{1}{c|}{rd} &
\multicolumn{1}{c|}{00001011} & PEQ.S \\
\cline{2-11}

&
\multicolumn{2}{|c|}{11010} &
\multicolumn{2}{c|}{10} &
\multicolumn{2}{c|}{rs2} &
\multicolumn{1}{c|}{rs1} &
\multicolumn{1}{c|}{000} &
\multicolumn{1}{c|}{rd} &
\multicolumn{1}{c|}{00001011} & PLT.S \\
\cline{2-11}

&
\multicolumn{2}{|c|}{11011} &
\multicolumn{2}{c|}{10} &
\multicolumn{2}{c|}{rs2} &
\multicolumn{1}{c|}{rs1} &
\multicolumn{1}{c|}{000} &
\multicolumn{1}{c|}{rd} &
\multicolumn{1}{c|}{00001011} & PLE.S \\
\cline{2-11}

\end{tabular}
\end{table*}

\myCommentDMQbis{
An important addition of the Xposit extension are the quire instructions. Since the quire is a single internal register of the \gls{PAU}, the instructions that operate with it do not have to specify a quire register number. For example, the quire clear instruction does not have any parameters. It is decoded and then executed internally by the PAU, which simply sets the quire register to 0. The quire fused operations only have to specify the posit registers of the two values that will be multiplied. Then, the accumulation is performed implicitly on the quire.
}

\section{Synthesis Results} \label{sec:synthesis_results}

In this section, we present the \gls{FPGA} and \gls{ASIC} synthesis results of PERCIVAL. The details of its \gls{PAU} and the IEEE 754 \gls{FPU} using 32 and 64-bit formats are also included. In this manner, the hardware cost of posit numbers and the quire are highlighted and compared with other implementations.

\subsection{FPGA Synthesis} \label{sec:fpga_synthesis}

The \gls{FPGA} synthesis was performed using Vivado v.2020.2 targeting a Genesys II (Xilinx Kintex-7 XC7K325T-2FFG900C) \gls{FPGA}. Different configurations of \gls{FPU} and \gls{PAU} were tested, the results of which are shown in Table~\ref{tab:fpga_synthesis_core}. Since the critical path does not traverse the arithmetic units of the core, in all of the cases the timing constraint of 20ns was met and the timing slack was +0.177ns.

\begin{table*}
\centering
\caption{Comparison of \gls{FPGA} synthesis results with different configurations of \gls{FPU}, marked as F and D for 32 and 64-bit numbers respectively, and 32-bit \gls{PAU} with quire.}
\label{tab:fpga_synthesis_core}
\resizebox{\textwidth}{!}{%
\begin{tabular}{@{}lcccccccc@{}}
\toprule
\multicolumn{1}{c}{} &
  \multicolumn{4}{c}{PAU} &
  \multicolumn{4}{c}{No PAU} \\ \midrule
\multicolumn{1}{c}{} &
  F &
  D &
  FD &
  - &
  F &
  D &
  FD &
  - \\ \midrule
\begin{tabular}[c]{@{}l@{}}Total core \\ (LUT, FF)\end{tabular} &
  (50318, 25727) &
  (55900, 27652) &
  (57129, 27996) &
  (44693, 23636) &
  (35402, 21618) &
  (40740, 23599) &
  (41260, 23945) &
  (28950, 19579) \\
\begin{tabular}[c]{@{}l@{}}FPU area\\ (LUT, FF)\end{tabular} &
  (3726, 1008) &
  (6352, 1905) &
  (7612, 2245) &
  - &
  (4046, 973) &
  (6626, 1905) &
  (8163, 2244) &
  - \\
\begin{tabular}[c]{@{}l@{}}PAU area\\ (LUT, FF)\end{tabular} &
  (11796, 2979) &
  (11810, 2979) &
  (11803, 2979) &
  (11879, 2985) &
  - &
  - &
  - &
  - \\ \bottomrule
\end{tabular}%
}
\end{table*}

The bare CVA6 without a \gls{FPU} or \gls{PAU} requires 28950 \glspl{LUT} and 19579 \glspl{FF}. Including support for 32-bit floating-point numbers increases the number of \glspl{LUT} and \glspl{FF} by 6452 and 2039 respectively. This difference grows to 12310 \glspl{LUT} and 4366 \glspl{FF} when using also the double-precision D extension. Note that these values are larger than simply the \gls{FPU} area, since they also include other elements such as the floating-point register file, instruction decoding and interconnections. These other non-\gls{FPU} elements require 2406 \glspl{LUT} and 1066 \glspl{FF} in the 32-bit case and 4147 \glspl{LUT} and 2122 \glspl{FF} in the 64-bit case.

Comparing the overall cost of including posit support with the cost of including IEEE floating-point support, a significant difference can be seen. Adding 32-bit posit operations and quire support to the CVA6 requires 15743 \glspl{LUT} and 4057 \glspl{FF}, which is comparable to the FD floating-point configuration. Out of this area, 3864 \glspl{LUT} and 1072 \glspl{FF} are occupied by the non-\gls{PAU} elements mentioned in the previous floating-point analysis.

The synthesis results reveal that the \gls{PAU} requires significantly more resources than the \gls{FPU} available in the CVA6. In particular, the 32-bit \gls{PAU} with quire occupies 2.94 times as many \glspl{LUT} and 3.07 times as many \glspl{FF} as the 32-bit \gls{FPU}. To better understand these results, in Table~\ref{tab:fpga_synthesis_pau} the area requirements of the different modules inside the \gls{PAU} are presented. The most interesting value shown in this table is the area occupied by the posit \gls{MAC} unit, which corresponds to almost half of the total area of the \gls{PAU}. 

When compared with the floating-point units, which do not include an accumulation register, the area requirements of the quire could be separated. Thus, the posit \gls{MAC} and the quire rounding to posit can be subtracted from the total \gls{PAU} area to obtain a value of 5326 \glspl{LUT} and 1312 \glspl{FF}. This outcome is now much closer to the synthesis results of the \gls{FPU}, as the \gls{PAU} without quire occupies 1.32 times as many \glspl{LUT} and 1.35 times as many \glspl{FF}. These results match previous works~\cite{ciocirlan2021Accuracy}, where authors also report an increase of around 30\% in \gls{FPGA} resources when comparing their 32-bit \gls{PAU} without quire with a 32-bit \gls{FPU}.

In our case, the actual value of not including a quire would be even smaller, as the cost of allocating the 512-bit quire in the \gls{PAU} and computing its 2's complement, which are included in the \gls{PAU} top, should also be subtracted. However, the synthesis tool does not include these details.

\begin{table}
\centering
\caption{\gls{FPGA} synthesis area results of the \gls{PAU} desegregated into its individual components.}
\label{tab:fpga_synthesis_pau}
\begin{tabular}{@{}l
                S[table-number-alignment = right, table-figures-decimal = 0]
                S[table-number-alignment = right, table-figures-decimal = 0]
                @{}}
\toprule
Name           & LUTs  & FFs  \\ \midrule
PAU top        & 593   & 1063 \\
Posit Add      & 784   & 106  \\
Posit Mult     & 736   & 73   \\
Posit ADiv     & 413   & 43   \\
Posit ASqrt    & 426   & 33   \\
Posit MAC      & 5644  & 1541 \\
Quire to Posit & 889   & 126  \\
Int to Posit   & 176   & 0    \\
Long to Posit  & 331   & 0    \\
ULong to Posit & 425   & 0    \\
Posit to Int   & 499   & 0    \\
Posit to Long  & 379   & 0    \\
Posit to UInt  & 228   & 0    \\
Posit to ULong & 358   & 0    \\ \midrule
PAU total      & 11879 & 2985 \\
PAU w/o quire    & 5346  & 1318 \\ \bottomrule
\end{tabular}
\end{table}

\subsection{ASIC Synthesis} \label{sec:asic_synthesis}

The 32-bit \gls{PAU} with quire and the 32-bit \gls{FPU} configuration present in PERCIVAL were synthesized targeting TSMC's 45nm standard-cell library to further study their hardware cost in \glspl{ASIC}. The synthesis was performed using Synopsys Design Compiler with a timing constraint of 5ns, which was met in both cases, and a toggle rate of 0.1.

The 32-bit \gls{FPU} within CVA6 requires an area of 30691~$\mu\text{m}^2$ and consumes 27.26~mW of power. On the other hand, the 32-bit \gls{PAU} with quire requires an area of 76970~$\mu\text{m}^2$ and consumes 67.73~mW of power. This follows the same trend shown in the \gls{FPGA} synthesis, as the \gls{PAU} with quire is significantly larger, 2.51x, and consumes more power, 2.48x, than the \gls{FPU}.

In addition, to better assess these values in comparison with other proposals, the \gls{PAU} available in CLARINET~\cite{sharma2021CLARINET} was also synthesized with the same parameters. We have chosen to evaluate this work because it integrates, to the best of our knowledge, the only other \gls{PAU} that contains a quire. In this case, the 32-bit \gls{PAU} with quire requires an area of 69920~$\mu$m$^2$ and consumes 68.31~mW of power. This is a decrease of around 10\% in area and a slight increase in power compared to our proposal, \myCommentDMQ{although ours implements a much larger set of posit functionality.}

Similarly as in Section~\ref{sec:fpga_synthesis}, the area and power results of the different elements inside the \gls{PAU} are presented in Table~\ref{tab:asic_synthesis_pau}. As can be seen, when subtracting the cost of the quire in the \gls{PAU}, the outcome is still higher than the 32-bit \gls{FPU}, but it becomes much closer. The 32-bit \gls{PAU} occupies 1.32 times as much area and consumes 1.38 times as much power as the 32-bit IEEE \gls{FPU} \myCommentDMQbis{FPNew~\cite{mach2021FPnew}}. \myCommentDMQ{However, it is noteworthy that some aspects of posit arithmetic are not yet fully studied. For example, most of the works presenting posit units have tackled the decoding and encoding phases using sign-magnitude. Nonetheless, more recent studies show that a 2's complement approach is more efficient~{\mbox{\cite{murillo2022Comparing}}}.}

\begin{table}
\centering
\caption{\gls{ASIC} synthesis area and power results of the 32-bit \gls{PAU} with quire desegregated into its individual components.}
\label{tab:asic_synthesis_pau}
\begin{tabular}{@{}l
                S[table-number-alignment = right, table-figures-decimal = 2]
                S[table-number-alignment = right, table-figures-decimal = 2]
                @{}}
\toprule
Name           & {Area ($\mu$m$^2$)} & {Power (mW)}\\ \midrule
PAU top        & 13462.15     & 12.69 \\
Posit Add      & 4075.31      & 3.59 \\
Posit Mult     & 8635.37      & 9.98 \\
Posit ADiv     & 2540.87      & 2.41 \\
Posit ASqrt    & 1722.84      & 1.61 \\
Posit MAC      & 30419.12     & 26.07 \\
Quire to Posit & 6026.76      & 4.04 \\
Int to Posit   & 905.99       & 0.68 \\
Long to Posit  & 1423.43      & 0.96 \\
UInt to Posit  & 869.77       & 0.66 \\
ULong to Posit & 1353.11      & 0.94 \\
Posit to Int   & 966.67       & 0.71 \\
Posit to Long  & 1810.33      & 1.38 \\
Posit to UInt  & 958.44       & 0.68 \\
Posit to ULong & 1800.22      & 1.33 \\ \midrule
PAU total      & 76970.38     & 67.73 \\
PAU w/o quire    & 40524.62     & 37.62 \\ \midrule
CLARINET PAU    & 69920.02     & 68.31 \\ \bottomrule
\end{tabular}
\end{table}

\section{Posit vs IEEE-754 Benchmarks} \label{sec:posit_vs_ieee}

One of the benefits of PERCIVAL is that an accurate and fair comparison can be made between posit and IEEE floating point. The main advantage of having support for native posit and IEEE floating point simultaneously on the same core is that identical benchmarks can be run on both number representations to compare them. In this work, we have chosen to benchmark the \acrfull{GEMM} and the max-pooling layer, used to down-sample the representation of neural networks. These examples showcase the use of the quire and posits both in the \gls{PAU} and in the \gls{ALU}, loading and storing from memory and leveraging the posit register file.

The \gls{GEMM} and max-pooling codes for posits and IEEE floats have been written in C, including inline assembly for the required posit and float instructions. The floating-point code has also been written in inline assembly to provide exactly the same sequence of instructions to the core. The \gls{GEMM} code for floats is shown in Figure~\ref{alg:float_gemm} and the analogous version for posits \myCommentDMQ{using the quire} is shown in Figure~\ref{alg:posit_gemm}. These codes have been compiled using the modified version of LLVM with the Xposit RISC\nobreakdash-V extension as specified in Section~\ref{sec:compiler_support}, and serve as an example of how this extension can be leveraged. Therefore, the final target architecture is RV64GCXposit. The \texttt{-O2} optimization flag has been used to obtain an optimized code in every case.

\begin{figure}
\begin{algorithmic}
    \REQUIRE Float matrices \texttt{a} and \texttt{b} of size \texttt{n}$\times$\texttt{n}.
    \ENSURE Float matrix \texttt{c} = \texttt{ab}.
    \FOR{i = 0 \TO n-1}
        \FOR{j = 0 \TO n-1}
            \STATE \textbf{asm}("fmv.w.x ft0,zero":::); \COMMENT{Set ft0 to 0}
            \FOR{k = 0 \TO n-1}
                \STATE\textbf{asm}
                    \STATE \quad "flw \qquad\ ft1,0(\%0)" \COMMENT{Load float a and b}
                    \STATE \quad "flw \qquad\ ft2,0(\%1)"
                    \STATE \quad "fmadd.s ft0,ft1,ft2,ft0" \COMMENT{Accumulate on ft0}
                    \STATE \quad :: "r" (\&a[i * n + k]), "r" (\&b[k * n + j]):
                \STATE\textbf{end asm}
            \ENDFOR
            \STATE\textbf{asm}
                \STATE \quad "fsw ft0,0(\%1)" \COMMENT{Store the result in c}
                \STATE \quad : "=rm" (c[i * n + j]) : "r" (\&c[i * n + j]):
            \STATE\textbf{end asm}
        \ENDFOR
    \ENDFOR
\end{algorithmic}
\caption{32-bit floating-point GEMM using the F RISC\protect\nobreakdash-V extension.}
\label{alg:float_gemm}
\end{figure}

\begin{figure}
\begin{algorithmic}
    \REQUIRE Posit matrices \texttt{a} and \texttt{b} of size \texttt{n}$\times$\texttt{n}.
    \ENSURE Posit matrix \texttt{c} = \texttt{ab}.
    \FOR{i = 0 \TO n-1}
        \FOR{j = 0 \TO n-1}
            \STATE \textbf{asm}("qclr.s":::); \COMMENT{Clear the quire}
            \FOR{k = 0 \TO n-1}
                \STATE\textbf{asm}
                    \STATE \quad "plw \qquad pt0,0(\%0)" \COMMENT{Load posit a and b}
                    \STATE \quad "plw \qquad pt1,0(\%1)"
                    \STATE \quad "qmadd.s pt0,pt1" \COMMENT{Accumulate on the quire}
                    \STATE \quad :: "r" (\&a[i * n + k]), "r" (\&b[k * n + j]):
                \STATE\textbf{end asm}
            \ENDFOR
            \STATE\textbf{asm}
                \STATE \quad "qround.s pt2" \COMMENT{Round the quire to a posit}
                \STATE \quad "psw \qquad pt2,0(\%1)" \COMMENT{Store the result in c}
                \STATE \quad : "=rm" (c[i * n + j]) : "r" (\&c[i * n + j]) :
            \STATE\textbf{end asm}
        \ENDFOR
    \ENDFOR
\end{algorithmic}
\caption{Posit GEMM using the Xposit RISC\protect\nobreakdash-V extension with the quire accumulator.}
\label{alg:posit_gemm}
\end{figure}

\subsection{Accuracy}

The accuracy differences between posits and floats are studied for the \gls{GEMM} benchmark. \myCommentDMQ{Furthermore, each arithmetic is executed with and without using fused {\gls{MAC}} operations, which in posit arithmetic include the quire. In the cases without quire or \texttt{FMADD}, each fused operation is substituted by a multiplication and an addition.} The results obtained using the 64-bit IEEE 754 format are considered the golden solution and used to compute the {\gls{MSE}} of the 32-bit posit and the 32-bit IEEE 754 floating point. In all cases, the inputs are square matrices with the same random values. \myCommentDMQ{These input values} are generated from a uniform distribution \myCommentDMQ{in intervals of the form $[-10^i, 10^i],\ i\in \{-1, 0, 1, 2, 3\}$. This results in 5 different sets of inputs. These intervals allow for a study of the impact of the input data range on the {\gls{GEMM}}.} These random values are generated as 64-bit IEEE 754 numbers and then converted to the two other formats with the aid of the SoftPosit~\cite{leong2020SoftPosit} library.

\begin{figure*}
    \centering
    \includegraphics[width=\textwidth]{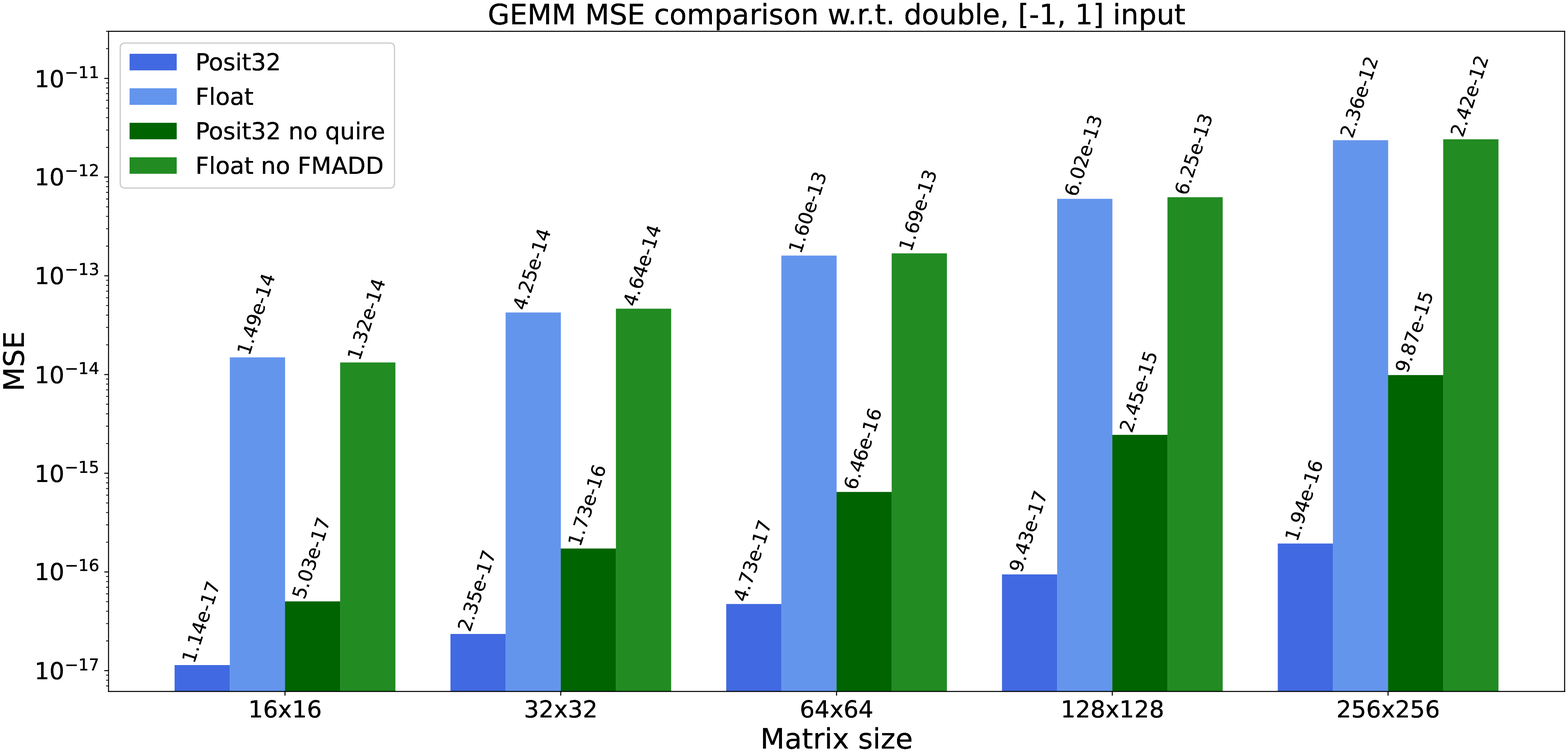}
    \caption{\myCommentDMQ{{\mbox{\gls{MSE}}} results of posits and floats with respect to doubles in the {\gls{GEMM}} test with input values in $[-1, 1]$. Note the logarithmic Y-axis. Blue (green) bars show the results with (without) fused \gls{MAC} and quire operations.}}
    \label{fig:mse}
\end{figure*}

The \gls{MSE} results are shown in Table~\ref{tab:gemm_mse} for different matrix sizes \myCommentDMQ{and input ranges. Additionally, Figure~{\ref{fig:mse}} shows the {\gls{MSE}}  in the $[-1, 1]$ case. We decided to give slightly more attention to this case since many applications normalize their values. As can be seen,} for $256\times 256$ matrices, the difference between \glspl{MSE} is around four orders of magnitude \myCommentDMQ{when using fused operations. This is reduced to two orders of magnitude if the quire is not used. Note that when using floats, the accuracy difference between employing fused \texttt{FMADD} operations or not is minimal.}

If we compare how \myCommentDMQ{the {\gls{MSE}}} scales when increasing the \myCommentDMQ{matrix} size, it can be seen that posit numbers present a better behavior thanks to the quire register. \myCommentDMQ{This is true in all ranges of input values. Overall, the impact of the quire is significant among all test cases, and its extra cost is justified by the results.}

\myCommentDMQ{These results go} in line with our previous work~\cite{murillo2021EnergyEfficient}, where a similar benchmark was performed using hardware simulations \myCommentDMQ{with an input interval of $[-2, 2]$}. The \gls{MSE} results on 32-bit floats and posits \myCommentDMQ{follow the same trends} given in Table~\ref{tab:gemm_mse}.

\begin{table*}
\centering
\caption{\myCommentDMQ{{\gls{GEMM}} {\gls{MSE}} comparison between IEEE 754 floating-point and posit numbers.}}
\label{tab:gemm_mse}
\begin{tabular}{@{}clccccc@{}}
\toprule
Input values                   & Matrix size & $16 \times 16$  & $32 \times 32$  & $64 \times 64$  & $128 \times 128$ & $256 \times 256$ \\ \midrule
\multirow{4}{*}{[-0.1, 0.1]}   & IEEE 754    & \num{1.385e-18} & \num{4.429e-18} & \num{1.523e-17} & \num{6.347e-17}  & \num{2.407e-16}  \\
 & Posit32           & \num{3.157e-21} & \num{6.110e-21} & \num{1.158e-20} & \num{2.014e-20} & \num{3.497e-20} \\
 & IEEE 754 no FMADD & \num{1.515e-18} & \num{4.752e-18} & \num{1.566e-17} & \num{6.524e-17} & \num{2.432e-16} \\
 & Posit32 no quire  & \num{2.146e-20} & \num{6.726e-20} & \num{2.371e-19} & \num{7.805e-19} & \num{2.203e-18} \\ \midrule
\multirow{4}{*}{[-1, 1]}       & IEEE 754    & \num{1.490e-14} & \num{4.251e-14} & \num{1.602e-13} & \num{6.019e-13}  & \num{2.361e-12}  \\
 & Posit32           & \num{1.138e-17} & \num{2.355e-17} & \num{4.729e-17} & \num{9.430e-17} & \num{1.937e-16} \\
 & IEEE 754 no FMADD & \num{1.324e-14} & \num{4.637e-14} & \num{1.686e-13} & \num{6.246e-13} & \num{2.416e-12} \\
 & Posit32 no quire  & \num{5.028e-17} & \num{1.727e-16} & \num{6.457e-16} & \num{2.447e-15} & \num{9.870e-15} \\ \midrule
\multirow{4}{*}{[-10, 10]}     & IEEE 754    & \num{1.371e-10} & \num{3.998e-10} & \num{1.581e-09} & \num{5.922e-09}  & \num{2.378e-08}  \\
 & Posit32           & \num{8.549e-13} & \num{1.475e-12} & \num{3.055e-12} & \num{6.355e-12} & \num{1.295e-11} \\
 & IEEE 754 no FMADD & \num{1.300e-10} & \num{4.304e-10} & \num{1.708e-09} & \num{6.026e-09} & \num{2.447e-08} \\
 & Posit32 no quire  & \num{3.878e-12} & \num{1.341e-11} & \num{7.500e-11} & \num{3.282e-10} & \num{1.41e-09}  \\ \midrule
\multirow{4}{*}{[-100, 100]}   & IEEE 754    & \num{1.412e-6}  & \num{4.206e-6}  & \num{1.544e-5}  & \num{6.402e-5}   & \num{2.405e-4}   \\
 & Posit32           & \num{4.819e-8}  & \num{8.266e-8}  & \num{1.760e-7}  & \num{6.150e-7}  & \num{1.506e-6}  \\
 & IEEE 754 no FMADD & \num{1.293e-6}  & \num{5.052e-6}  & \num{1.595e-5}  & \num{6.503e-5}  & \num{2.440e-4}  \\
 & Posit32 no quire  & \num{3.077e-7}  & \num{1.230e-6}  & \num{4.295e-6}  & \num{2.804e-5}  & \num{1.569e-4}  \\ \midrule
\multirow{4}{*}{[-1000, 1000]} & IEEE 754    & \num{1.503e-2}  & \num{3.936e-2}  & \num{1.509e-1}  & \num{6.069e-1}   & \num{2.391}      \\
 & Posit32           & \num{5.293e-3}  & \num{8.573e-3}  & \num{1.900e-2}  & \num{3.746e-2}  & \num{8.265e-2}  \\
 & IEEE 754 no FMADD & \num{1.675e-2}  & \num{4.815e-2}  & \num{1.644e-1}  & \num{6.323e-1}  & \num{2.433}     \\
 & Posit32 no quire  & \num{4.168e-2}  & \num{1.570e-1}  & \num{5.669e-1}  & \num{2.365}     & \num{9.586}     \\ \bottomrule
\end{tabular}
\end{table*}

\myCommentDMQ{When removing the quire, posits still have a lower MSE than floats except in the $[-1000, 1000]$ case. This can be explained by posit's tapered precision. When the numbers' exponents are closer to $0$, they end up in the so-called ``golden zone'' of posits~{\mbox{\cite{dedinechin2019Posits}}}. This is the area where posits have more accuracy bits than floats thanks to their variable-length fields. However, when the accumulated values are large or very small, IEEE floats gain an advantage over posits without quire.}

\myCommentDMQ{
Particularly, this ``golden zone'' comprises values roughly in the interval $[10^{-6}, 10^6]$. In the test with input values in $[-1000, 1000]$, the absolute value of the final outputs averages $1.2\times10^6$ in the $16\times16$ matrix and $4.3\times10^6$ in the $256\times256$ case. As a comparison, even in the $256\times256$ multiplication, the $[-100, 100]$ input range only averages $4.3\times10^4$.
}

\subsection{Performance}

Besides the synthesis data presented in Section~\ref{sec:synthesis_results}, the execution time is a critical metric to study the hardware performance of posits and floats. The test has been performed executing the same \gls{GEMM} and max-pooling described previously on PERCIVAL, avoiding cold misses and averaging over 10 executions to obtain more accurate measurements.

\begin{table*}
\centering
\caption{\myCommentDMQ{GEMM timing comparison between IEEE 754 floating-point and posit numbers.}}
\label{tab:gemm_time}
\begin{tabular}{@{}lccccc@{}}
\toprule
Matrix size &
  \multicolumn{1}{l}{$16 \times 16$} &
  \multicolumn{1}{l}{$32 \times 32$} &
  \multicolumn{1}{l}{$64 \times 64$} &
  \multicolumn{1}{l}{$128 \times 128$} &
  \multicolumn{1}{l}{$256 \times 256$} \\ \midrule
32-bit float                 & 0.978 ms & 6.58 ms  & 52.1 ms  & 1.48 s & 13.9 s \\
64-bit float                 & 0.920 ms & 6.64 ms  & 69.4 ms  & 1.74 s & 15.0 s \\
Posit32                      & 0.949 ms & 7.30 ms  & 57.7 ms  & 1.48 s & 13.9 s \\
32-bit float no FMADD        & 1.16 ms & 8.69 ms  & 68.6 ms  & 1.61 s & 15.0 s \\
64-bit float no FMADD        & 1.26 ms & 9.36 ms  & 92.6 ms  & 1.92 s & 16.7 s \\
Posit32 no quire             & 1.27 ms & 9.63 ms  & 69.1 ms  & 1.61 s & 15.0 s \\
VividSparks Posit32 no quire & 7.95 ms  & 48.9 ms  & 345 ms   & 2.63 s & 21.1 s \\ \bottomrule
\end{tabular}
\end{table*}

\myCommentDMQ{The range of the input values does not affect performance. Thus, the values shown in Table~{\ref{tab:gemm_time}} for {\gls{GEMM}} are an average of the timings obtained in the 5 cases described previously. This gives a total of 50 executions in the {\gls{GEMM}} operation. In this case, when using fused {\gls{MAC}} operations and the quire,} the execution time of 32-bit posits is practically the same as that of single-precision floats for the larger matrix sizes, where the overhead execution of the extra qround.s instruction becomes negligible (see Figure~\ref{alg:posit_gemm}). This instruction is executed in the order of $\mathcal{O}(n^2)$ times, compared with the $\mathcal{O}(n^3)$ running time of the algorithm. This cost is noticeable for smaller values of $n$, when 32-bit posits are slightly slower than 32-bit and 64-bit floats. However, for larger matrix sizes, which are common in scientific applications and \glspl{DNN}, 32-bit posits perform equally as 32-bit floats and outperform 64-bit floats, since these instructions require more clock cycles to compute. Furthermore, as seen in the previous accuracy benchmark, 32-bit posits are orders of magnitude more accurate than 32-bit floats when performing this calculation. Therefore, they provide an alternative solution for the execution of kernels that make use of the dot product.

\myCommentDMQ{The quire and fused {\gls{MAC}} operations have a positive impact on timing performance. This is true in all test cases. Again, this performance increase stems from the extra clock cycles needed for a multiplication + an addition in comparison to only one fused operation.}

Additionally, for the sake of completeness, we have performed the same \gls{GEMM} timing test on a commercial core with support for posit arithmetic. RacEr is a GPGPU \gls{FPGA} provided by VividSparks that supports computation with Posit32 but does not include quire support\myCommentDMQ{, so its accuracy results are the same as the Posit32 no quire case}. It has 512 CPUs running at 300MHz with 32GB of DDR4 RAM. Table~{\ref{tab:gemm_time}} also includes the results of the \gls{GEMM} benchmark on this platform. As can be seen, our proposal provides significantly faster results than this commercial accelerator.

Regarding the max-pooling layers, three different configurations have been tested following common \glspl{DNN}. In LeNet-5, the input of this layer is 28x28x6, the pooling kernel is 2x2 and is applied with a stride of 2, creating a 14x14x6 output representation. In AlexNet, the input size is 54x54x96, the kernel size is 3x3 and is applied with a stride of 2, generating an output of size 26x26x96. Finally, ResNet-50 is the largest configuration we have tested, as its input is 112x112x64, the pooling kernel is 3x3 and again is applied with a stride of 2, creating a 55x55x64 output representation.

The results of executing these layers on PERCIVAL using the 32 and 64-bit IEEE floating-point and Posit32 representations are shown in Table~\ref{tab:maxpool_time}. Results show that 32-bit posits perform as fast as 32-bit floats but without the need for extra hardware, as the posit maximum operation is carried out reusing the integer \gls{ALU}. Double-precision floats are slower than 32-bit posits and floats by a factor of 1.4-1.7$\times$ due to the latency difference in the units as seen in the \gls{GEMM} benchmark.

\begin{table}
\centering
\caption{Max-pooling timing comparison between IEEE 754 floating-point and posit numbers.}
\label{tab:maxpool_time}
\begin{tabular}{@{}lccc@{}}
\toprule
Max-pooling layer & 32-bit float & 64-bit float & Posit32 \\ \midrule
LeNet-5 \quad(28x28x6) & 0.715ms      & 1.211ms      & 0.688ms \\
AlexNet \ \ \ (54x54x96) & 0.115ms      & 0.160ms      & 0.116ms \\
ResNet-50 (112x112x64) & 0.337ms      & 0.470ms      & 0.340ms \\ \bottomrule
\end{tabular}
\end{table}

 \section{Conclusions} \label{sec:conclusions}

This paper has presented PERCIVAL, an extension of the application-level CVA6 RISC\nobreakdash-V core, including all 32-bit posit instructions as well as the quire fused operations. These capabilities, integrated into a \acrlong{PAU} together with a posit register file, are natively incorporated while preserving IEEE 754 single- and double-precision floats.

Furthermore, the RISC\nobreakdash-V ISA has been extended with Xposit, which includes support for all posit and quire instructions. This allows the compilation and execution on PERCIVAL of application-level programs that make use of posits and floats simultaneously. To the best of our knowledge, this is the first work that enables complete posit and quire capabilities in hardware.

Synthesis results show that half the area dedicated to the \gls{PAU} is occupied by the quire and its operations. When comparing with the only previous work which includes quire capabilities~\cite{sharma2021CLARINET}, our proposal consumes slightly less power and only 10\% more area, while also providing full posit operations support. When focusing on the 32-bit \gls{PAU} without the quire, our proposal requires 32\% more area and 38\% more power than the 32-bit \gls{FPU}. This goes in line with the results of recent works which reuse the F RISC\nobreakdash-V extension~\cite{ciocirlan2021Accuracy}, where authors obtain a 30\% increase in \gls{FPGA} resources when comparing their \gls{PAU} to the \gls{FPU}.

The Posit vs IEEE-754 comparison benchmark results show that 32-bit posits are up to 4 orders of magnitude more accurate than 32-bit floats when calculating the \gls{GEMM} due to the quire. Moreover, they do not show a performance degradation compared with floats, thus providing a potential alternative when operating with real numbers. In addition, our proposal performs significantly better than available commercial solutions, obtaining up to \myCommentDMQ{8}$\times$ speedup when multiplying small matrices.

\myCommentDMQbis{
Some known limitations occur in the use of the quire. As it is a single internal register in the \gls{PAU}, PERCIVAL cannot support parallel accumulation into different independent accumulators. This also prevents safe automatic context switches, as the value of the quire cannot be loaded or stored in memory. Therefore, when developing programs for PERCIVAL this must be taken into account to not overwrite the value of the quire.
}

As future work, we plan to implement and evaluate on PERCIVAL large-scale scientific applications which make use of dot products, leveraging the accuracy gains of fused operations.


%



\ifCLASSOPTIONcompsoc
  \section*{Acknowledgments}
\else
  \section*{Acknowledgment}
\fi

This work was supported by a 2020 Leonardo Grant for Researchers and Cultural Creators, from BBVA Foundation, whose id is PR2003\_20/01, by the EU(FEDER) and the Spanish MINECO under grant RTI2018-093684-B-I00, and by the CM under grant S2018/TCS-4423.

\ifCLASSOPTIONcaptionsoff
  \newpage
\fi



\bibliographystyle{IEEEtran}
\bibliography{references}
%

%
\begin{IEEEbiography}[{\includegraphics[width=1in,height=1.25in,clip,keepaspectratio]{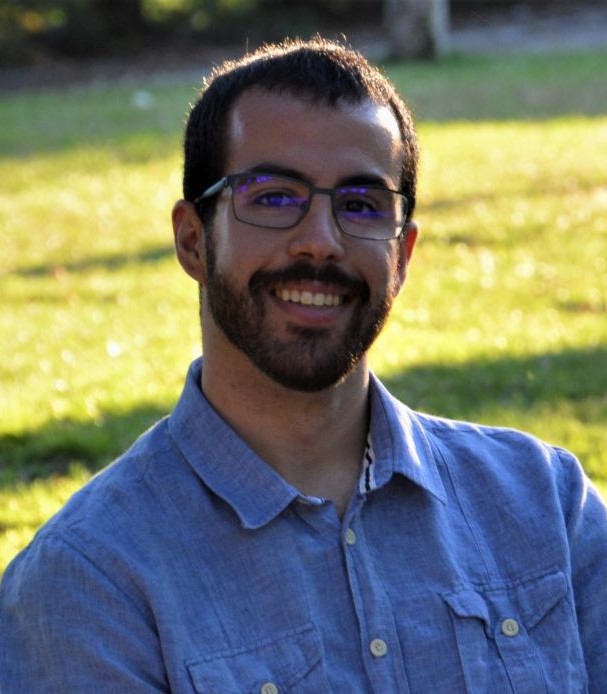}}]{David Mallasén}
David Mallasén Quintana received a BSc Degree in Computer Science and a BSc Degree in Mathematics in 2020 from the Complutense University of Madrid (UCM). From 2020 to 2022 he obtained a MSc Degree in Embedded Systems at KTH Royal Institute of Technology, specializing in embedded platforms. Currently he is pursuing a Ph.D. in Computer Engineering at UCM. His main research areas include computer arithmetic, computer architecture, embedded systems, and high-performance computing.
\end{IEEEbiography}

\begin{IEEEbiography}[{\includegraphics[width=1in,height=1.25in,clip,keepaspectratio]{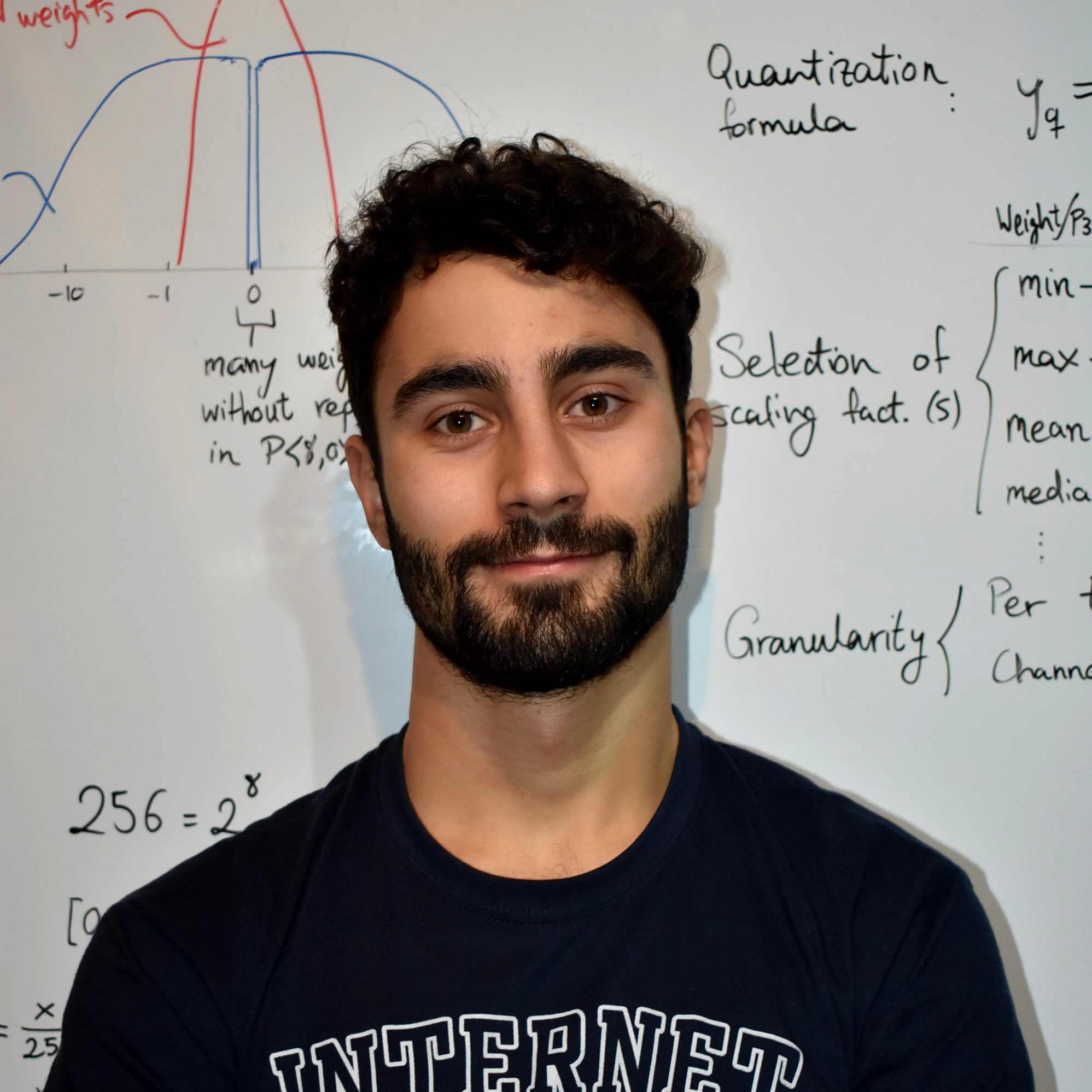}}]{Raul Murillo}
Raul Murillo studied Mathematics and Computer Science at Complutense University of Madrid (UCM), Spain, where he also received a MSc Degree in Computer Science in 2021. His main research interests include Approximate Computing, new Computer Arithmetic, and Deep Neural Networks (DNNs). He is currently pursuing a Ph.D. at UCM related to the previously mentioned areas.
\end{IEEEbiography}

\begin{IEEEbiography}[{\includegraphics[width=1in,height=1.25in,clip,keepaspectratio]{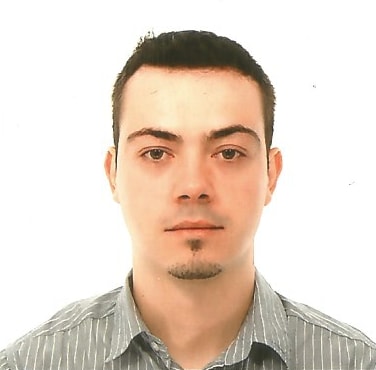}}]{Alberto A. Del Barrio}
Alberto A. Del Barrio received the Ph.D. degree in Computer Science from the Complutense University of Madrid (UCM), Madrid, Spain, in 2011. He has performed stays at Northwestern University, University of California at Irvine and University of California at Los Angeles. Since 2021, he is an Associate Professor (tenure-track, civil-servant) of Computer Science with the Department of Computer Architecture and System Engineering, UCM. His main research interests include Design Automation, Arithmetic and their application to the field of Artificial Intelligence. He is leading the PARNASO project, funded by the Leonardo Grants program by Fundación BBVA. The main objective is to natively integrate the posit format in a hardware/software platform. Since 2019 he is an IEEE Senior Member and since December 2020 he is an ACM Senior Member, too.
\end{IEEEbiography}

\begin{IEEEbiography}[{\includegraphics[width=1in,height=1.25in,clip,keepaspectratio]{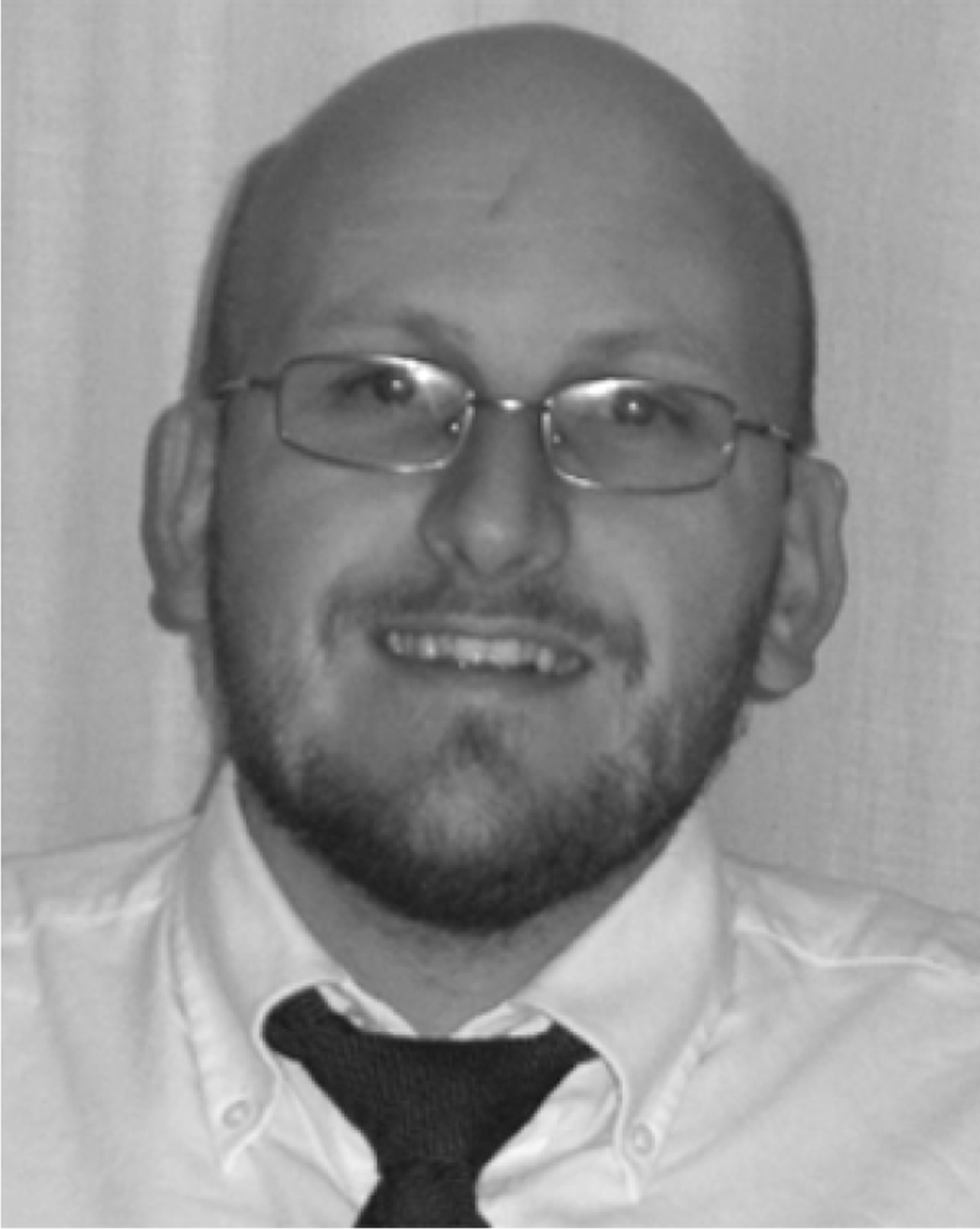}}]{Guillermo Botella}
Guillermo Botella received the M.A.Sc. degree in Physics (Fundamental) in 1998, the M.A.Sc. degree in Electronic Engineering in 2001 and the Ph.D. degree (Computer Engineering) in 2007, all from the University of Granada, Spain. He was a research fellow funded by EU working at University of Granada, Spain and the Vision Research Laboratory at University College London, UK. After that, he joined as Assistant Professor at the Department of Computer Architecture and Automation of Complutense University of Madrid, Spain where he is currently Associate Professor. He has performed research stays acting also as visiting professor from 2008 to 2012 at the Department of Electrical and Computer Engineering, Florida State University, Tallahassee, USA. His current research interests include Image and Video Processing for VLSI, FPGAs, GPGPUs, Embedded Systems, and novel computing paradigms such as analog and quantum computing. Since 2019 he has become an IEEE Senior Member.
\end{IEEEbiography}

\vfill

\begin{IEEEbiography}[{\includegraphics[width=1in,height=1.25in,clip,keepaspectratio]{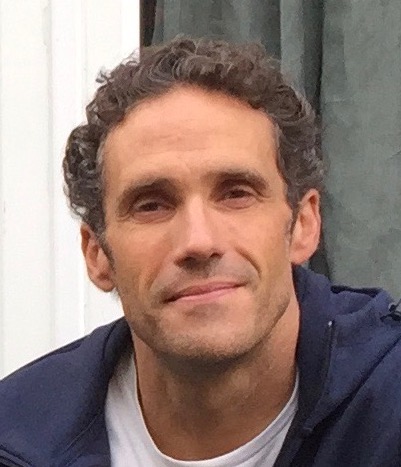}}]{Luis Piñuel}
Luis Piñuel is an Associate Professor of the Department of Computer Architecture and Systems Engineering at the Universidad Complutense de Madrid, Spain.  He received his M. Sc. and Ph.D. degrees in Computer Science from the Universidad Complutense de Madrid (UCM) in 1996 and 2003, respectively. His research interests include computer architecture, high-performance computing, embedded systems, and resource management for emerging computing systems. In these fields, he is co-author of more than 70 publications in prestigious journals and international conferences, several book chapters and he has advised or co-advised 5 PhD dissertations. Worried about improving knowledge transfer between research institutions and industry, he has directed more than 12 research contracts with different companies (Texas Instruments, Imagination Technologies, Indra, ...). He has also served as evaluator for several national agencies and has also been member of the Board of Directors of the Spanish Computer Architecture Society (SARTECO).
\end{IEEEbiography}

\begin{IEEEbiography}[{\includegraphics[width=1in,height=1.25in,clip,keepaspectratio]{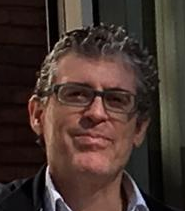}}]{Manuel Prieto-Matias}
Manuel Prieto Matias obtained a Ph.D. degree from Complutense University of Madrid (UCM) in 2000.  Since 2002, he has been a Professor at the Department of Computer Architecture at UCM, being a Full Professor since 2019. His research interests include high-performance computing, non-volatile memory technologies, accelerators, and code generation and optimization. His current focus is on effectively managing resources on emerging computing platforms, emphasizing the interaction between the system software and the underlying architecture. Manuel has co-authored over 100 scientific publications in journals and conferences in parallel computing and computer architecture. He is a member of the ACM.
\end{IEEEbiography}





\vfill


\end{document}